\documentclass[a4paper]{ceurart}

\usepackage{booktabs}
\usepackage{tabularx}
\usepackage{array}
\usepackage{hyperref}
\usepackage{microtype}
\usepackage{graphicx}
\usepackage{amsmath}
\usepackage[T1]{fontenc}
\usepackage[utf8]{inputenc}
\usepackage{pdflscape}
\definecolor{customgreen}{RGB}{0,176,80}
\newcolumntype{P}[1]{>{\raggedright\arraybackslash}p{#1}}

\copyrightyear{2026}
\copyrightclause{Copyright for this paper by its authors.
  Use permitted under Creative Commons License Attribution 4.0
  International (CC BY 4.0).}
\conference{ALIT4ALL 2026: Workshop on AI Literacy for All,
  co-located with AIED 2026}

\begin{document}

 \title{Programming Language Policy as an AI Literacy Equity Problem: 
A 15-Nation Comparative Analysis}

\author[1]{Adrian-Marius Dumitran}[%
  orcid=0009-0005-3547-5772,
  email=marius.dumitran@unibuc.ro,
]
\cormark[1]
\fnmark[1]
\author[1]{Iulia-Maria Popescu}
\fnmark[1]

\address[1]{University of Bucharest, Faculty of Mathematics and
  Computer Science, Bucharest, Romania}

\cortext[1]{Corresponding author.}
\fntext[1]{These authors contributed equally.}
\begin{abstract}
The promise of AI literacy ``for all'' confronts a structural 
challenge embedded in how nations organise secondary computer 
science education. In most systems, a general-track subject --- 
Digital Literacy, ICT, TIC, or SNT --- bears the weight of 
universal AI literacy, while a specialist Informatics course 
serves STEM pathways separately. Yet the content and depth of 
the general track are shaped by governance decisions made largely 
with reference to the specialist one. This paper presents a 
comparative analysis of curricula and examination frameworks 
across fifteen countries, identifying two structural challenges. 
First, in several systems a significant portion of students 
completes secondary education without any formal programming 
exposure. Second, among those who do receive CS education, a \emph{Syntax Ceiling} emerges: Python-based instruction reaches most students, while the algorithmic depth associated with C++ remains concentrated in elite STEM tracks. Drawing on reform cases spanning 
centralised mandates (France, China, Japan), assessment-driven 
systems (Poland, Romania, South Korea), and recent universal 
reforms (Switzerland, Kazakhstan), we show that governance 
structures and high-stakes examinations are the primary drivers 
of both challenges --- and that specialist and general-track 
language choices are rarely independent, linked through shared 
teacher pipelines that curriculum policy seldom acknowledges. 
Achieving genuine AI literacy for all requires confronting not 
just curriculum content, but the access architectures and 
resource constraints that determine who receives it --- and at 
what depth.
\end{abstract}


\maketitle


\section{Introduction}
\label{sec:intro}
The global expansion of artificial intelligence has elevated
programming from a specialist skill to a foundational literacy.
Governments, international bodies, and educators increasingly argue
that every student --- not just future engineers --- needs sufficient
computational understanding to participate meaningfully in an
AI-driven society. UNESCO's Guidance for Generative AI in
Education~\cite{UNESCO2023} and numerous national AI strategies
reflect this consensus. In most national systems, this mandate falls
primarily on general-track subjects --- variously called Digital
Literacy, ICT, TIC, or SNT --- designed for all students regardless
of pathway, while specialist Informatics or CS courses serve
STEM-track students separately. It is the general-track subject that
bears the weight of the ``for all'' promise. Yet that promise runs
into a structural obstacle: in many systems, access to programming
education is not universal but gated by track selection, geography,
and socioeconomic circumstance.

This paper argues that secondary CS education is undergoing a
consequential bifurcation. A broad, Python-based layer is emerging
as a near-universal AI literacy foundation, while a narrower,
C++-based layer persists as the gateway to elite algorithmic tracks.
We term this divide the \emph{Syntax Ceiling}: the point at which
the depth of computational education becomes inaccessible to students
outside specialist pathways. Crucially, it is not simply a
pedagogical preference --- it is a governance outcome, shaped by
curriculum mandates, national examination structures, and the
washback effects they generate~\cite{Alderson1993,Messick1996}. Its
consequence for AI literacy is direct: students above the ceiling are
positioned --- by the depth of their computational instruction --- to
understand, evaluate, and potentially contribute to the design of AI
systems, while students below it receive instruction oriented toward
AI use and application. This maps onto the four-tier competency
hierarchy of~\cite{Ng2021}, which separates \emph{knowing and
understanding} and \emph{using and applying} AI from the more
demanding tier of \emph{evaluating and creating} it. We treat the
strength of this link as a tension to be examined --- it is actively
contested among practitioners --- rather than a settled equivalence.

For designers of AI literacy programs and curricula, this
bifurcation creates a practical challenge. Students arriving in
general-track AI literacy courses do so with computational
backgrounds shaped by governance decisions made years earlier --- in
examination boards and curriculum working groups --- often with
little reference to AI literacy goals. This paper focuses on upper
secondary education (ISCED~3), where formal programming mandates and
high-stakes examinations crystallise these governance effects most
clearly. AI literacy exposure can occur earlier, embedded in lower
secondary mathematics or science subjects, as Norway's LK20
curriculum~\cite{Udir2020} illustrates; such pre-secondary pathways
complement but fall outside this study's scope.

We present a comparative analysis of fifteen national systems. Our
contribution is threefold: (1) we frame programming language
stratification as an AI literacy equity problem, connecting curriculum
policy to the ``for all'' imperative; (2) we demonstrate that
governance structures and assessment mechanisms are the primary
drivers of both access gaps and depth stratification; and (3) we
analyze concrete reform cases --- from Switzerland's MAV transition
to Kazakhstan's AI-first integration --- showing that both barriers
persist even among these leading systems, suggesting they are
structural rather than incidental.

\section{Related Work}
\label{sec:related}

AI literacy is broadly defined as the set of competencies enabling
individuals to critically evaluate, interact with, and ethically use
AI systems~\cite{Long2020}. While frameworks vary, most include some
notion of understanding how AI works --- which in practice implicates
programming knowledge, particularly for understanding data-driven
systems. The question of which students receive this knowledge, and
at what depth, is therefore central to any equity-focused AI literacy
agenda.

Yet AI literacy frameworks rarely specify the programming depth
required to operationalize their competencies. Long and
Magerko~\cite{Long2020} identify seventeen competencies spanning
perception, ethics, and interaction with AI --- but stop short of
prescribing whether meaningful engagement with AI requires the
ability to read Python, to write it, or to understand the
lower-level computational logic that underlies it. This ambiguity
matters for curriculum design: a student who can adjust a
pre-trained model's parameters via a high-level API occupies a
very different epistemic position from one who can implement a
sorting algorithm or understand memory allocation. The former
represents AI literacy as \emph{use}; the latter as
\emph{comprehension}. Both are legitimate goals, but they correspond
to different instructional requirements, different teacher
preparation pipelines, and --- as this paper demonstrates ---
different governance structures. For AI literacy program designers,
the practical question is therefore not only whether students have
been exposed to programming, but which layer of the computational
stack that exposure reached.

Large-scale evidence on language selection is most robust at the
tertiary level. Mason et al.'s 2024 Global Survey~\cite{Mason2024}
highlights a convergence toward dominant languages and significant
institutional inertia in CS1/CS2 curricula, documenting a ``clear
joint lead'' between Python and Java, with both utilized by 46\% of
surveyed institutions. This parity reflects a strategic bifurcation:
Python broadens the incoming student funnel while Java maintains
alignment with industrial software engineering standards. Philip
Guo's 2014 systematic audit of top-tier US departments found that
80\% had shifted to Python~\cite{Guo2014}, a shift that historically
sets downstream secondary standards.

Comparative K--12 research has historically focused on the structural
positioning of informatics rather than specific language policies.
Reports by Eurydice~\cite{Eurydice2022} and Webb et
al.~\cite{Webb2017} document whether CS is compulsory or integrated,
highlighting challenges in teacher capacity and curriculum sequencing
--- but rarely treat programming languages as primary objects of
comparison, and rarely distinguish between what general-track and
specialist-track students receive. The ``washback
effect''~\cite{Alderson1993,Messick1996} --- whereby high-stakes
examinations constrain classroom practice regardless of formal
curriculum flexibility --- has been documented extensively in
language testing but remains underexplored in CS education. Our work
extends this line of inquiry to secondary systems across 15
countries, treating language policy, track differentiation, and
their equity implications as primary objects of analysis.

Rather than assume a link between programming exposure and AI
literacy, we ground it in how competency frameworks define advanced
AI skills. In the four-tier hierarchy of~\cite{Ng2021} (introduced
above), the \emph{evaluate and create} tier --- genuine critical
agency over AI --- presupposes the ability to read, trace, and modify
a system's computational logic, not merely operate its interface;
work mapping AI literacy across educational levels similarly finds
that such higher-order skills require computational depth beyond
introductory programming~\cite{Chee2025}. We therefore do not claim
that a specific language makes a student AI-literate, nor that C++
alone produces AI architects. We argue that the \emph{depth} of
computational instruction --- shaped by governance decisions about
language, curriculum scope, and track access --- is a structural
condition that opens or closes access to the higher-order competencies
those frameworks identify as most consequential.

\section{Methodology and Analytical Framework}
\label{sec:framework}

\subsection{Country Selection and Data Sources}

The fifteen national systems analysed in this paper were selected
through purposive sampling to represent each governance archetype
and a range of mandate levels, while ensuring geopolitical and
developmental diversity. Selection criteria prioritised: (1)
structural influence --- systems whose policy decisions
demonstrably shape regional or global secondary CS norms; (2)
reform activity --- systems that have undergone documented
transitions in programming language policy or mandate level since
2010; and (3) documentation availability --- systems with
publicly accessible statutory curricula, examination
specifications, or ministry decrees in a language amenable to
analysis. This third criterion introduces a known constraint:
systems without institutionalised documentation --- including many
lower-income nations across Africa, South and Southeast Asia, and
small island states --- could not be included. Their absence from
this cohort does not imply the absence of CS education or AI
literacy efforts, but is an artefact of the document-based
methodology this study employs.

Data collection drew on three source types: statutory instruments 
(federal ordinances, ministry decrees, and national curriculum 
frameworks), assessment specifications (examination board syllabi 
and accepted language lists), and official national education 
portals. Where documentation was not available in English, 
translated versions were consulted; terminological distinctions 
between ``ICT Literacy'' and ``Informatics'' were treated with 
particular care, as these carry significant classification 
implications. Federal and regionally devolved systems present 
inherent classification challenges, as governance models and 
language choices may vary substantially across sub-national 
boundaries; our classifications reflect the dominant or 
nationally representative pattern where a single authoritative 
standard does not exist.

A known limitation of this document-based approach is the
\emph{implementation fidelity gap}: statutory frameworks and
examination specifications define the intended curriculum, but
classroom reality is mediated by teacher preparation, resource
availability, and institutional inertia. Practitioners in several
systems report significant divergence between mandated frameworks
and enacted practice, particularly where specialist CS teacher
pipelines remain underdeveloped. This paper analyses governance
mechanisms and intended curricula; the distance between policy
and classroom practice is an important complementary research
agenda outside this study's scope.

\subsection{Analytical Framework}

We classify national systems along two dimensions. The first is
\emph{Mandate Level}: whether programming is elective,
track-mandated (available to some students), or universal
(required of all). The second is \emph{Governance Archetype}:
the institutional mechanism through which language choices are
made and enforced. We identify four archetypes, summarised in
Table~\ref{tab:archetypes}. The archetypes were constructed
inductively around one question --- where the effective authority
over the classroom programming language resides: with teachers and
schools, a central ministry, a national exam's accepted-language
list, or sub-national authorities.

\begin{table*}[ht]
\caption{Governance Archetypes and Language Outcomes}
\label{tab:archetypes}
\footnotesize
\begin{tabular}{P{2.6cm} P{4.2cm} P{3.8cm} P{2.8cm}}
\toprule
\textbf{Archetype} & \textbf{Description} & \textbf{Language Outcome} & \textbf{Examples} \\
\midrule
Model A: Sovereign-Led
  & Ministry decree defines language and curriculum
  & Python (mass) / C++ (elite) by decree
  & France, China, UAE, Japan \\
\addlinespace
Model B: Assessment-Driven
  & National exam implicitly mandates language list via washback
  & Python + C++ via exam washback effect
  & Poland, Romania, Finland, South Korea \\
\addlinespace
Model C1: Decentralized (regional exams)
  & Sub-national authority; regional exams re-centralize through entrance requirements
  & Language varies; OOP bias in university entrance exams
  & UK, Germany \\
\addlinespace
Model C2: Fully Decentralized
  & Teacher/school autonomy; no national CS exam
  & Python dominant via professional norms and industry standards
  & Switzerland (pre-2024), Norway, Netherlands \\
\bottomrule
\end{tabular}
\end{table*}

A critical finding from applying this framework is that governance
archetypes do not always produce their stated outcomes. In Model
C2 systems (fully decentralised), teacher autonomy is
theoretically preserved, but professional norms and university
entrance expectations often produce \emph{de facto}
standardisation --- a phenomenon we term \emph{assessment
lock-in}. In Model C1 systems, such as the UK and Germany,
awarding bodies and regional Abitur requirements act as
re-centralising forces, producing language convergence toward
Python and Java even in the absence of a central ministry decree.
Each country was then coded on two features --- whether a central
authority fixes the language for assessment, and whether curriculum
flexibility is formally preserved --- and assigned to an archetype
by the \emph{primary mechanism driving language outcomes} rather
than the formal structure alone. This is what separates Model~B
(exam washback produces a de facto mandate despite formal
language-agnosticism) from Model~A (an explicit decree). These
classifications involve interpretive judgement, and the boundaries
are not always sharp.

These two dimensions --- mandate level and governance archetype
--- jointly determine two distinct equity problems: whether
students access programming at all (the \emph{access gap}), and
whether they reach algorithmic depth or only surface-level
literacy (the \emph{Syntax Ceiling}). Critically, in many
systems these two tracks are delivered by the same teachers,
creating a \emph{resource coupling} between general-track and
specialist-track language choices that operates below the level
of formal policy and that curriculum designers rarely account
for.

\section{Access and Depth: Two Challenges for AI Literacy For All}
\label{sec:barriers}

\subsection{The Access Gap}
\label{sec:access}

For the ``for all'' promise of AI literacy to be meaningful,
the general-track subject --- TIC, SNT, Digital Literacy, or
its local equivalent --- must actually reach every student.
In a significant number of systems, however, the most
fundamental barrier to AI literacy is not the depth of
instruction but its existence. Track selection --- often
determined at age 14 or earlier, and correlated with
socioeconomic background --- determines whether a student
encounters programming at all. As the Eurydice comparative
analysis documents, programming instruction across European
secondary systems ranges from compulsory standalone subjects
to entirely optional or cross-subject integration, meaning
that in several systems a student may complete general
secondary education with digital tool literacy but no formal
programming exposure~\cite{Eurydice2022}. Germany illustrates
this concretely: with curriculum authority devolved to 16
Länder and no federal informatics mandate, a student's
exposure depends heavily on school type, state of residence,
and track --- leaving large portions of the non-gymnasium
population with no guaranteed AI literacy pathway at all.
Switzerland's ongoing MAV reform represents a deliberate
policy response to precisely this kind of fragmentation,
establishing Informatics as a mandatory subject across all
26 cantons from 2024. Yet even this ambitious reform operates exclusively within 
the gymnasium pathway, attended by fewer than one quarter 
of Swiss youth~\cite{BFS2024}; the majority who enter the 
vocational dual-track system before the gymnasium stage fall 
outside its scope. A mandate that reaches one in
four students is a meaningful advance --- but it is not
``for all''.

\subsection{The Python--C++ Trade-off}
\label{sec:tradeoff}

Among students who do receive CS education, a second tension
emerges around the choice of primary programming language.
Across most systems in our cohort this manifests as a
Python--C++ divide: Python favoured for its readability,
accessibility, and alignment with AI and data science
workflows; C++ for its computational depth, relevance to
competitive programming, and applicability to
performance-critical domains such as cybersecurity and
systems programming. In some systems --- notably the UK and
parts of Germany --- Java occupies the specialist-track role
with comparable structural rigour. The equity concern is not
that Python is insufficient for AI work --- modern AI
development is overwhelmingly Python-first --- but that
general-track Python instruction rarely delivers the
algorithmic depth that specialist tracks achieve, regardless
of which language carries it.

The choice is not cost-free in either direction. Python
across both tracks gains accessibility and teacher coherence
but risks omitting the lower-level computational reasoning
that underpins AI systems at the infrastructure level.
Preserving C++ in the specialist track maintains that depth
for a minority but imposes a Python catch-up burden, since
virtually all major AI frameworks are Python-first. Systems
resolve this tension differently --- China mandates the split
explicitly~\cite{ChinaAI2017}; France achieves broad Python
access without eliminating specialist depth~\cite{France2019};
Poland has organically displaced C++ through exam task
redesign~\cite{CKE2023} --- and Section~5 examines these and
further cases in detail.
\section{Comparative Evidence}
\label{sec:evidence}

Table~\ref{tab:profiles} summarises the access and depth profile of 
fifteen national systems across four governance archetypes. Together they illustrate a global landscape 
in transition: from systems where programming remains the preserve 
of specialist tracks, to recent universal reforms designed explicitly 
around AI literacy goals. The subsections below examine the primary 
reform archetypes in detail, with attention to the governance 
mechanisms that drove each transition and their implications for 
equitable AI literacy access.

\begin{table}[ht]
\caption{Access and Depth Profiles Across 15 National Systems}
\label{tab:profiles}
\footnotesize
\begin{tabular}{P{1.8cm} P{1.6cm} P{3.4cm} P{3.5cm} P{3.6cm}}
\toprule
\textbf{Country} & \textbf{Mandate Level} & \textbf{General Track}
  & \textbf{STEM / Informatics Track} & \textbf{AI Literacy Reach} \\
\midrule
France       & Universal        & Python (SNT)
  & Python/Java (NSI)       & Broad --- all students \\
\addlinespace
China        & Bifurcated       & Python (mass AI mandate)
  & C++ (competitive track) & Wide but stratified \\
\addlinespace
Japan        & Universal        & Python (Info I)
  & Python + DNCL (exam decoupled) & Broad --- assessment decoupled \\
\addlinespace
UAE          & Universal        &   AI literacy mandatory K--12 
  & Python/Java (advanced)  & Broad --- centralised reform \\
\addlinespace
Poland       & Track-mandated   & Python (Matura data tasks)
  & C++ (competitive, retained) & Medium --- exam-driven pivot \\
\addlinespace
Romania & Track-mandated 
  & ICT literacy (TIC); AI literacy pilot from 2025--26
  & C++/Pascal (Bac.); Python transition planned 2030
  & Low $\to$ Broad (planned 2026); implementation uncertain \\

\addlinespace
South Korea  & Universal        & Python (SW mandate 2018)
  & C/C++ (specialised schools) & Broad --- AI integrated \\
\addlinespace
Finland      & Track-mandated   & Python/Java (LOPS 2021)
  & Python/Java (IT elective) & Medium --- curriculum-embedded \\
\addlinespace
UK           & Track-mandated   & Python (post-2014 reform)
  & Python/Java (A-level CS) & Medium --- awarding body driven \\
\addlinespace
Germany      & Track-mandated   & Python/Java (by Länder)
  & C++/Java (Abitur)       & Medium --- federal fragmentation \\
\addlinespace
Switzerland  & Universal (2026) & Python (MAV reform)
  & Python/Java (ETH guidelines) & Medium --- gymnasium track (~23\% of youth)
 \\
\addlinespace
Norway       & Integrated       & Python/JS (LK20)
  & Python/Java (IT elective) & Broad --- cross-subject \\
\addlinespace
Netherlands  & Track-mandated   & Python (2019 reform)
  & Python/Java/C\# (modular) & Medium --- elective structure \\
\addlinespace
Bulgaria     & Elective         & IT literacy (no AI literacy) 
  & C++ (Informatics Track) & Narrow --- specialist only \\
\addlinespace
Kazakhstan   & Universal        & Python (2025/26 AI integration)
  & C++ (NIS pipeline)      & Broad --- AI-first mandate \\
\bottomrule
\end{tabular}
\end{table}

\subsection{Reform Archetypes: From Sovereign Decrees to Federal
            Escalation}
\label{sec:reform}

\subsubsection{State-Led Mandates: France, China, and the UAE.}
In highly centralized systems (Model A), transitions are defined by a
``structural break'' enforced via ministry decree. France's 2019
reform replaced mathematical pseudocode with a unified Python
standard for the mandatory SNT and specialized NSI
tracks~\cite{France2019}. This bifurcation ensures that all students
receive Python-based AI literacy exposure, while the NSI track
provides algorithmic depth for those pursuing STEM. The governance
mechanism is direct: a central decree specifies the language, the
curriculum, and the assessment criteria simultaneously.

China's approach is more explicitly stratified. The 2017 AI
Development Plan~\cite{ChinaAI2017} established a categorical
mandate: Python for mass AI literacy across all general secondary
students, C++ strictly reserved for the competitive programming track
and the National Olympiad in Informatics pipeline. This dual-track
architecture is explicit policy, designed to produce both a broad
base of AI-literate citizens and a narrow elite of world-class
algorithm designers. From an equity perspective, this model is effective at closing 
the access gap while making the Python--C++ stratification 
an explicit instrument of national policy rather than an 
incidental outcome.

The UAE's 2025 Cabinet-approved mandate represents the 
most recent and ambitious instance of sovereign-led reform 
in this cohort: AI literacy is introduced as a compulsory 
subject from kindergarten through Grade 12 across all 
public schools, integrated within the Computing, Creative 
Design and Innovation subject without extending school 
hours~\cite{UAE2025}. Unlike France and China, the UAE 
mandate is explicitly AI-first --- it does not mandate a 
programming language but instead frames computational 
thinking and AI ethics as the primary literacy goals.

\subsubsection{Japan's Decoupled Reform.}
Japan's 2022 Information I curriculum reform represents a distinctive
institutional response to the assessment-language alignment problem.
Python is standardized for classroom instruction across all secondary
schools~\cite{MEXT2018,Kanemune2022}, while a domain-specific
pseudocode (DNCL) is used for the national university entrance
examination~\cite{DNC2024}. This decoupling strategy ensures that no
student is disadvantaged by their specific Python implementation or
IDE choice during high-stakes assessment, while simultaneously
delivering universal Python-based instruction. The approach addresses
a tension common to Model B systems: the washback effect tends to
freeze whatever language appears in the exam. By using a pseudocode
for assessment, Japan preserves the pedagogical benefits of Python
while neutralizing the assessment lock-in problem.

\subsubsection{Assessment-Driven Governance: Poland and Romania.}
Poland and Romania illustrate the power of assessment-driven
governance (Model B) to shape AI literacy outcomes --- for better
and worse. In both systems, the national examination (Matura and
Baccalaureate respectively) maintains a closed list of accepted
languages, creating a functional mandate even when the curriculum
text is formally language-agnostic. Teachers align instruction with
the exam list to ensure student success, effectively removing
pedagogical freedom --- a textbook washback
effect~\cite{Alderson1993,Messick1996}.

In Poland, this has produced a Python pivot for the majority of
students as Matura exam tasks shifted toward data processing and
library-based programming~\cite{CKE2023}. Python has become the
dominant submission language, displacing C++ as the practical
choice for most candidates --- a shift driven by exam task
redesign rather than curricular intent. The Polish core
curriculum~\cite{MEN2017} had formally remained
language-agnostic throughout this transition, yet the outcome is
a rapid, system-wide gain in AI-relevant programming literacy
produced entirely through assessment design. This demonstrates
that in Model B systems, exam design is \emph{de facto}
curriculum policy for AI literacy --- and that policymakers
who wish to shift language practice without revising curriculum
text have a direct and underutilised lever available to them.

Romania's Baccalaureate maintains a stronger C++ presence in
the exam list~\cite{Romania2023}, producing a different depth
profile from the same governance archetype. Algorithmic rigour
is preserved in the specialist track, while the general track
has historically been limited to legacy ICT literacy with no
programming or AI content. Two parallel reforms are now
underway: the TIC curriculum, redesigned to include AI
literacy content, will reach all students entering 9th grade
from the 2026--2027 academic year, moving the general track
from near-zero AI literacy exposure toward universal coverage
phased in through the cohort by 2030; separately, the
Baccalaureate language list targets a transition from
C++/Pascal to Python as the sole examination language by
2030. Whether both transitions will be implemented as planned
depends critically on teacher preparation pipelines that
operate on longer timescales than policy documents.

\subsubsection{Switzerland's Federal Escalation.}
Switzerland serves as the study's primary reform case for closing
the access gap at scale. Historically operating under Model C2
with highly variable CS exposure across its 26 cantons,
Switzerland is completing a transition to a nationwide mandate
through the Maturity Recognition Ordinance (MAV) reforms. By
2026, Informatics is recognised as a mandatory foundational
subject across all cantons --- a structural shift from
decentralised variation to federal coordination. Python and Java
have emerged as the primary instructional languages, reflecting
the influence of ETH Z\"{u}rich's academic guidelines rather
than a federal language decree; the MAV reform specifies the
subject as mandatory but leaves language choice to cantons and
institutions within that framework.

This case demonstrates that even highly decentralised systems
can achieve coordinated reform through federal policy --- and
that the access gap can be substantially reduced when
institutional anchors provide curricular coherence across
sub-national boundaries. It also illustrates a structural
limit that ambitious reforms can leave unaddressed: the MAV
applies exclusively to the gymnasium pathway, attended by
approximately one fourth of Swiss youth. The majority who enter
the vocational dual-track system fall outside its scope,
meaning that the access gap is closed for the academic
minority while remaining largely intact for the vocational
majority.

\subsubsection{Kazakhstan's AI-First Integration.}
Kazakhstan represents a reform archetype characterised by
late adoption and strategic reframing. Rather than inheriting
a legacy CS curriculum built around algorithmic depth,
Kazakhstan's 2025/2026 reforms integrate AI literacy directly
into the foundational secondary curriculum, establishing
Python as the primary language for all students while
specialised NIS network schools maintain a C++ pipeline for
competitive programming and engineering tracks. This
dual-layer structure is structurally similar to China's
categorical mandate~\cite{ChinaAI2017}, with the distinction
that Kazakhstan is adopting it as a system-building measure
rather than a reform of an established CS tradition.

\subsection{Transition Mechanisms and AI Literacy Implications}
\label{sec:transitions}

Table~\ref{tab:transitions} summarises the primary transition
mechanisms for seven reform-active systems in our cohort,
mapping the shift from previous standards to current ones and
identifying the specific AI literacy implications of each
transition type.

\begin{table*}[ht]
\caption{Programming Language Transition Mechanisms and AI Literacy Implications}
\label{tab:transitions}
\footnotesize
\begin{tabular}{P{1.8cm} P{2.2cm} P{3.2cm} P{3.2cm} P{4.2cm}}
\toprule
\textbf{Country} & \textbf{Transition Type} & \textbf{Previous Standard}
  & \textbf{Current Standard} & \textbf{AI Literacy Implication} \\
\midrule
France
  & Structural reform
  & Math pseudocode (AlgoBox)
  & Python mandate (SNT/NSI, 2019)
  & Universal Python foundation for AI tools; depth in NSI \\
\addlinespace
Japan
  & Decoupled reform
  & ICT ethics/office focus (pre-2022)
  & Python (Info I - 2022) + DNCL exam (2025)
  & Broad AI literacy; exam neutrality via pseudocode \\
\addlinespace
China
  & Categorical mandate
  & Pascal/C++ for Olympiads only
  & Python (mass) / C++ (elite), 2017
  & Explicit stratification: mass literacy vs. programming depth \\
\addlinespace
Poland
  & Assessment-led pivot
  & C++ dominant (pre-2015 Matura)
  & Python dominant via data tasks (2023)
  & Rapid AI literacy gain via exam design reform \\
\addlinespace
Switzerland
  & Federal escalation
  & Decentralised / canton-dependent
  & Python/Java, MAV reform (2024--2026)
  & Closes access gap for gymnasium track; coordinates policy across cantons \\
\addlinespace
South Korea
  & SW mandate reform
  & ICT literacy, specialized schools
  & Python AI mandate, all schools (2018)
  & National AI readiness; C++ maintained for specialists \\
\addlinespace
Kazakhstan
  & AI-first integration
  & ICT literacy, legacy languages
  & Python + AI curriculum (2025-2026)
  & AI-first integration: Python foundation 
    without legacy C++ detour \\
\bottomrule
\end{tabular}
\end{table*}

The table reveals three broad transition strategies with
distinct AI literacy implications. Decree-based transitions
(France, China) produce clear, policy-specified outcomes but
require strong central governance capacity. Assessment-led
transitions (Poland) are less predictable but can produce
rapid AI literacy gains as a side-effect of exam task
redesign --- without requiring any change to curriculum text.
Federal escalation (Switzerland) demonstrates that
decentralised systems can achieve coordinated reform, but
require extended timelines and strong institutional anchors
to bridge sub-national variation.

\section{Discussion and Implications}
\label{sec:discussion}

The reform trajectories documented in this study serve a dual
purpose. Analytically, they validate the two-dimensional framework
introduced in Section~3: governance archetype and mandate level
jointly determine both who accesses programming education and what
depth they receive. Practically, they constitute a structured
inventory of transition mechanisms available to curriculum designers
and governance bodies navigating language policy decisions. France's
structural break required strong central decree capacity; Poland's
Python pivot required only exam task redesign; Switzerland's federal
escalation required an institutional anchor to coordinate across
decentralised cantons. No single mechanism transfers wholesale across
governance contexts, but the mapping of mechanism to archetype
offered here allows policymakers to identify the levers most likely
to operate within their own system's constraints.

\subsection{Syntax Stratification as a Global Trend}
\label{sec:stratification}

Our findings suggest that the global CS curriculum is not converging
on a single language but is instead undergoing a process of
\emph{syntax stratification}. Python has achieved near-universal
presence as the language of general-track AI literacy instruction,
while C++ --- and in some systems Java --- persists as the language
of elite algorithmic tracks across most of the 15-country cohort.
This bifurcation is not accidental: it is the product of governance
decisions made at the national examination level, reinforced by
university entrance expectations and labour market signals.

We clarify what this does and does not imply. Differentiated
pathways --- where some students receive deeper specialist
instruction --- are a normal feature of secondary systems and are
not inherently inequitable. The concern is narrower: it arises when
access to the deeper pathway is systematically correlated with
socioeconomic background, geography, or track decisions made before
students have meaningful agency, and when the general track offers no
later route toward higher-order competencies. The Syntax Ceiling is
not a problem because general-track students learn Python rather than
C++; it is a problem when Python instruction is resourced in ways that
foreclose the \emph{evaluate and create} competencies~\cite{Ng2021}
--- and when those foreclosed are disproportionately the already
disadvantaged.

The equity implications extend directly to AI system design. As AI
tools become embedded across professional domains --- medicine, law,
engineering, public administration --- the question of who can
evaluate, modify, or architect these systems is increasingly
determined by who received algorithmic depth in secondary education.
The digital divide of 2026 is thus shifting: it is no longer defined
by simple access to coding, but by the divergence between students
oriented toward AI use and those with access to the depth needed to
evaluate and build AI systems~\cite{Ng2021}. Crucially, syntax
stratification is not an inevitable outcome of differentiated CS
education. France's SNT/NSI structure demonstrates that universal
Python access and specialist algorithmic depth can coexist within a
single system. China's categorical mandate shows the stratification
can be made explicit policy rather than left as an incidental outcome
of governance inertia. The question for AI literacy program designers
is therefore not whether stratification exists, but whether it is
deliberately designed, transparently communicated, and equitably
structured.

\subsection{The Dominance of the Washback Effect}
\label{sec:washback}

The study confirms that the national examination is the most powerful
tool of governance for AI literacy outcomes. Whether through Model A
mandates or Model B constraints, the high-stakes assessment (Matura,
Baccalaureate, Gaokao, A-level) dictates classroom reality. In the
UK, where awarding bodies such as OCR and AQA define the accepted
language list for A-level Computer Science~\cite{DfE2013}, Python has
become dominant not through ministerial decree but through exam board
standardization --- a textbook case of market-driven re-centralization
within a nominally decentralized system. South Korea's 2018 Software
Education mandate~\cite{SouthKorea2022} similarly demonstrates that
even Model B systems can achieve broad AI literacy outcomes when exam
design and curriculum reform are aligned. This has a critical
implication for AI literacy policy: in most systems, reforming the
curriculum text without reforming the examination is insufficient.
Poland's Python pivot was produced by exam task redesign, not
curricular intent.

\subsection{Policy Implications for AI Literacy Equity}
\label{sec:policy}

Our analysis yields four findings for AI literacy policy.

\textbf{First}, the access gap and the Python--C++ trade-off are
distinct problems needing distinct responses. Expanding mandate level
addresses access but not depth; France shows universal access can
coexist with differentiated depth, while improving elite tracks does
nothing for students who never encounter programming. Both must be
addressed --- and in sequence, since depth stratification matters
only once access is substantially closed.

\textbf{Second}, national examinations are the most powerful lever,
particularly in Model B systems. Which languages are accepted, what
task types appear, what skills are assessed --- these shape classroom
practice more immediately than curriculum text. Intentional exam
reform toward Python and AI-relevant tasks (data analysis, model
evaluation, prompt engineering) can produce rapid, system-wide shifts
without curriculum revision, as Poland demonstrates.

\textbf{Third}, which students receive algorithmic depth --- and which
receive only surface-level Python literacy --- has equity implications
beyond programming skill. By differentiating students on the depth
they receive, current tracking architectures risk reproducing
socioeconomic inequalities at the level of computational agency.
Addressing this needs not just curriculum reform but reform of
tracking and pathway-selection mechanisms, many decided before
secondary school and thus beyond the reach of CS education policy
alone.

\textbf{Fourth}, language choices for specialist and general-track
subjects are rarely independent in practice. Where specialist CS
teacher pipelines are thin --- the majority of our cohort --- the same
instructor often delivers both the general-track digital literacy
subject and the specialist Informatics course. This \emph{teacher
resource coupling} means the specialist track's language constrains
what is feasible in the general track, and vice versa. General-track
AI literacy curricula therefore cannot be designed as pedagogically
independent of the specialist track: in resource-constrained systems,
the specialist language arrives in the general classroom through the
teacher, whatever the curriculum document says.

\section{Limitations}
\label{sec:limitations}
Several limitations should be acknowledged. First, the document-based
methodology analyses intended curricula and statutory frameworks, not
enacted classroom practice; the \emph{implementation fidelity gap}
between policy and delivery is likely significant where specialist CS
teacher pipelines are thin, so our findings describe governance
structures and their designed outcomes rather than lived classroom
experience. Second, the cohort is a purposive sample biased toward
systems with strong documentation and reform activity; many
lower-income nations are absent --- an artefact of method, not
evidence that challenges are absent there, where governance
constraints are likely more acute. Third, we focus on upper secondary
(ISCED~3); earlier exposure embedded in lower-secondary mathematics or
science (e.g.\ Norway's LK20) and informal pathways fall outside our
scope. Fourth, both AI technology and education policy evolve quickly,
so some Table~2 classifications --- notably Kazakhstan's 2025/2026 and
Romania's planned 2030 reforms --- may already be in transition; the
framework is an analytical tool, not a static snapshot.

\section{Conclusions and Future Work}
\label{sec:conclusions}
This paper examined two structural challenges to equitable AI literacy
in secondary education: the access gap that excludes many students from
programming entirely, and the Python--C++ trade-off that shapes the
depth those who do receive CS education develop --- and thus how far
they reach toward the higher-order, ``evaluate and create'' end of AI
literacy. Both are governance outcomes, driven by mandate structures
and examinations more than by pedagogy.

Across our cohort the pattern is clear: Python is the near-universal
general-track language, while C++ persists in elite algorithmic tracks
--- a bifurcation produced at the examination level and reinforced by
university entrance and labour-market signals. The ICT-to-AI-literacy
transition is underway, but unevenly. Even where shared standards now
emerge --- notably the 2026 OECD/European Commission framework~\cite{OECD2026}
--- realisation stays fragmented, because governance and examinations,
not frameworks, determine what reaches the classroom. The reform cases
illustrate the available levers: France's decree shows universal access
and specialist depth can coexist; Poland shows exam redesign moves
faster than curriculum revision; Switzerland shows even decentralised
systems can coordinate, though the gymnasium scope of the MAV bounds
its reach; Kazakhstan shows late adopters can build CS education around
AI literacy from the outset.

Four directions remain. \emph{Teacher preparation}: one instructor
often delivers both tracks, and the C++-to-Python retraining gap is a
poorly documented bottleneck. \emph{The AI-first question}: whether
integrating AI literacy before traditional programming yields
sufficient computational reasoning is an open empirical question.
\emph{Tracking}: which populations are routed into specialist versus
general tracks, and what interventions at pathway selection can address
it. \emph{Excluded systems}: selected partly for reform activity, our
cohort under-represents low-reform, low-access settings --- often those
most relevant to ``for all,'' and the priority for extending this work.

\begin{acknowledgments}
The authors used LLMs including Claude (Anthropic), ChatGPT (OpenAI),
and Gemini (Google) to assist with manuscript revision and improving
text clarity. All content was verified and approved by the authors,
who take full responsibility for the accuracy and integrity of this
work.
\end{acknowledgments}



\begin{thebibliography}{21}

\bibitem{Alderson1993}
J.~C. Alderson and D.~Wall,
``Does washback exist?''
\emph{Applied Linguistics}, vol.~14, no.~2, pp.~115--129, 1993.
\url{https://doi.org/10.1093/applin/14.2.115}

\bibitem{Messick1996}
S.~Messick,
``Validity and washback in language testing,''
\emph{Language Testing}, vol.~13, no.~3, pp.~241--256, 1996.
\url{https://doi.org/10.1177/026553229601300302}

\bibitem{Long2020}
D.~Long and B.~Magerko,
``What is AI literacy? Competencies and design considerations,''
in \emph{Proc.\ 2020 CHI}, 2020.
\url{https://doi.org/10.1145/3313831.3376727}

\bibitem{Ng2021}
D.~T.~K. Ng, J.~K.~L. Leung, S.~K.~W. Chu, and M.~S. Qiao,
``Conceptualizing AI literacy: An exploratory review,''
\emph{Computers and Education: Artificial Intelligence}, vol.~2,
p.~100041, 2021.
\url{https://doi.org/10.1016/j.caeai.2021.100041}

\bibitem{Chee2025}
K.~N. Chee, B.~Lim, and S.~Y. Tan,
``A competency framework for AI literacy: Variations by different
learner groups and an implied learning pathway,''
\emph{British Journal of Educational Technology}, 2025.
\url{https://doi.org/10.1111/bjet.13556}

\bibitem{Mason2024}
R.~Mason, Simon, B.~A. Becker, T.~Crick, and J.~H. Davenport,
``A global survey of introductory programming courses,''
in SIGCSE 2024.
\url{https://doi.org/10.1145/3626252.3630827}

\bibitem{Guo2014}
P.~Guo,
``Python is now the most popular introductory language at top U.S.\
universities,''
\emph{Communications of the ACM}, 2014.
\url{https://cacm.acm.org/blogs/blog-cacm/176450}

\bibitem{Eurydice2022}
Eurydice,
\emph{Informatics Education at School in Europe}.
Publications Office of the European Union, 2022.
\url{https://doi.org/10.2797/268406}

\bibitem{Webb2017}
M.~Webb et al.,
``Computer science in K--12 school curricula of the 21st century:
Issues, influencing factors and exemplary countries,''
\emph{Education and Information Technologies}, vol.~22, no.~2,
pp.~403--468, 2017.
\url{https://link.springer.com/article/10.1007/s10639-016-9493-x}

\bibitem{ChinaAI2017}
State Council of China,
\emph{New Generation Artificial Intelligence Development Plan}.
Beijing: State Council, 2017.
\url{http://www.gov.cn/zhengce/content/2017-07/20/content_5211996.htm}

\bibitem{CKE2023}
Central Examination Commission (CKE),
\emph{Annual Report on the Matura Exam in Informatics}.
Warsaw: CKE, 2023.
\url{https://cke.gov.pl/}

\bibitem{France2019}
Minist\`ere de l'\'Education Nationale,
``Programme de sciences num\'eriques et technologie (SNT) de seconde
g\'en\'erale et technologique,'' 2019.
\url{https://www.education.gouv.fr/bo/19/Special1/MENE1901633A.htm}

\bibitem{DNC2024}
National Center for University Entrance Examinations (DNC),
``Common test for university admissions: Information I and DNCL,''
2024.
\url{https://www.dnc.ac.jp/albums/abm.php?d=4&f=abm00005032.pdf}

\bibitem{UNESCO2023}
UNESCO,
\emph{Guidance for Generative AI in Education and Research}.
Paris: UNESCO, 2023.
\url{https://doi.org/10.54675/PCSP7350}

\bibitem{MEXT2018}
Ministry of Education, Culture, Sports, Science and Technology
(MEXT),
``Revision of the course of study for upper secondary schools,''
2018.
\url{https://www.mext.go.jp/en/policy/education/elsec/title02/detail02/1373859.htm}

\bibitem{MEN2017}
Ministry of National Education (MEN),
\emph{Core Curriculum for General High School (Podstawa programowa
kszta\l{}cenia og\'olnego dla liceum og\'olnokszta\l{}c\k{a}cego)}.
Warsaw, 2017.
\url{https://www.gov.pl/web/edukacja-i-nauka}

\bibitem{Romania2023}
Ministerul Educa\c{t}iei,
``Repere metodologice pentru aplicarea curriculumului la clasa a
XI-a \^in anul \c{s}colar 2023--2024: Informatic\u{a} \c{s}i TIC.''
Bucharest: CNC, 2023.
\url{https://www.edu.ro/}

\bibitem{DfE2013}
Department for Education,
``National curriculum in England: Computing programmes of study,''
2013.
\url{https://www.gov.uk/government/publications/national-curriculum-in-england-computing-programmes-of-study}

\bibitem{Kanemune2022}
S.~Kanemune et al.,
``Informatics education in Japanese primary and secondary schools,''
Information Processing Society of Japan (IPSJ), 2022.
\url{https://ioi.te.lv/conf/c11_Kanemune.pdf}

\bibitem{SouthKorea2022}
Ministry of Education (South Korea),
``Policies on digital education and AI,'' 2022.
\url{http://english.moe.go.kr/}

\bibitem{Udir2020}
Utdanningsdirektoratet,
``L\ae{}replan i informasjonsteknologi (INF01-03),'' 2020.
\url{https://www.udir.no/lk20/inf01-03}

\bibitem{BFS2024}
Swiss Federal Statistical Office (BFS),
``Quote der Erstabschl\"{u}sse auf der Sekundarstufe~II
(Graduation Rates at Upper Secondary Level),''
Bern: BFS, 2024.
\url{https://www.bfs.admin.ch/bfs/de/home/statistiken/bildung-wissenschaft/uebertritte-verlaeufe-bildungsbereich/abschlussquoten.html}

\bibitem{UAE2025}
Ministry of Education (UAE),
``Artificial Intelligence introduced as mandatory subject
across all stages of government education,''
announced by H.H.\ Sheikh Mohammed bin Rashid Al Maktoum,
May 2025.
\url{https://www.thenationalnews.com/news/uae/2025/05/04/sheikh-mohammed-announces-introduction-of-ai-as-curriculum-subject-in-uae-schools/}

\bibitem{OECD2026}
OECD and European Commission,
\emph{Empowering Learners for the Age of AI: An AI Literacy Framework
for Primary and Secondary Education}.
Paris: OECD Publishing, 2026.
\url{https://doi.org/10.1787/65cd27d4-en}

\end{thebibliography}
\end{document}